# Institutional Trust and the Domestic AI Advantage: Evidence from DeepSeek and ChatGPT Users in China

Jiashen Huang[a*], Yu Jia[b], Xu Pan[c,d]

[a]*School of Journalism and Communication, Wuhan University, Wuhan, P.R. China;*

[b]*Center for Studies of Media Development, Wuhan University, Wuhan, P.R. China;*

[c]*School of Electrical and Electronic Engineering, Nanyang Technological University, Singapore;*

[d]*CFAR and IHPC, Agency for Science, Technology and Research (A*STAR), Singapore*

*Corresponding author: Jiashen Huang. School of Journalism and Communication, Wuhan University. Address: 299 Bayi Road, Wuchang District, Wuhan City, Hubei Province, P. R. China 430072. Email: jiashenhuang@whu.edu.cn

# Institutional Trust and the Domestic AI Advantage: Evidence from DeepSeek and ChatGPT Users in China

Public trust in generative artificial intelligence exhibits increasingly divergent patterns across national contexts, yet prevailing research largely overlooks the macro-structural forces underlying this divergence. This study argues that trust in AI is not merely a technical response to performance but a product of institutional refraction. We propose an 'Institutional Prism' framework to demonstrate how institutional trust shapes user trust in domestic (DeepSeek) and global (ChatGPT) large language models. Drawing on Cognitive-Affective Trust Theory, we distinguish between cognitive and affective dimensions of trust and analyze survey data from 405 Chinese users. The findings show that higher institutional trust is positively associated with stronger affective trust in domestic AI models and shifts cognitive evaluations in a more favorable direction. While under lower institutional trust, this domestic advantage weakens. These findings reveal that institutional trust has emerged as a core dimension of AI trust formation. By linking micro-level psychological judgments with macro-level governance, this research contributes a new perspective to human-machine communication.

## Introduction

The rapid proliferation of large language models (LLMs) has transformed artificial intelligence from a specialized tool into a primary infrastructure for everyday communication, information seeking, and decision-making. Prominent examples such as ChatGPT and DeepSeek have quickly entered public discourse, sparking renewed attention to how trust in AI is socially and culturally constructed.

In this study, we define AI trust as a multidimensional construct comprising cognitive trust and affective trust. The former refers to rational assessment of a system's competence and reliability, while the latter refers to an emotional bond and perceived benevolence of the technology (McAllister, 1995). Current research has predominantly focused on techno-centric factors, such as spanning system performance, interface design (e.g., anthropomorphism), and transparency mechanisms (Hancock et al., 2011). While these studies offer valuable insights into the psychological mechanics of trust, they often treat the user and the machine as analytically isolated from broader institutional contexts, overlooking the macro forces that govern technology adoption. Recent comparative scholarship has increasingly recognized a 'domestic AI advantage,' wherein localized computational systems often elicit greater baseline user trust owing to linguistic alignment and cultural resonance (Shen et al., 2026).From a macro-governance perspective, this alignment is a structural manifestation of 'values shaping' and 'values signaling' embedded within the sovereign architecture of AI systems (Koenig & Rone, 2026). In the context, generative AI models do not operate as tools; rather, they are regarded as specific institutional and normative regularities that reflect the sovereign state's regulatory visions.

However, a significant theoretical gap persists. Existing literature has yet to systematically examine how institutional trust moderates the relationship between an AI model's origin and a user's multidimensional trust. Institutional trust, defined as an individual's confidence in the integrity, competence, and regulatory capacity of overarching governance systems (Zucker, 1986). We still lack empirical clarity on how these macro-level situational forces interface with micro-level algorithmic preferences or aversion (Prahl & Van

Swol, 2017), and how this process varies across different user groups within the same national population.

While DeepSeek and ChatGPT differ in technological capabilities, their trust-building mechanisms are shaped by an interplay of functional performance and institutional legitimacy (Fosch-Villaronga & Millard, 2019). Furthermore, the competition for artificial intelligence leadership extends far beyond technological confrontation into the realm of ideological and geopolitical negotiation (Cheng & Zeng, 2023). Although computational systems are often perceived as objective and neutral infrastructure, they are inherently embedded with distinct socio-political values derived from state-specific regulatory visions. Therefore, we introduce the term 'Institutional Prism' to describe the moderating role of macro-level governance in shaping micro-level technological perceptions. Drawing on Pierre Bourdieu's (2014) concept of field refraction, we suggest that the national institutional environment acts as a lens that bends the perception of AI attributes to align with domestic social orders and political legitimacies.

This study integrates McAllister's (1995) Cognitive-Affective Trust Theory (CATT) with the 'Institutional Prism' framework. The aim of this paper is to investigate how the origin of an AI model (domestic vs. global) influences cognitive and affective trust, and to demonstrate how institutional trust serves as the refraction mechanism shaping these evaluations. By comparing DeepSeek and ChatGPT in the Chinese context, we argue that institutional trust functions as a vital social resource that channels state legitimacy into a relative trust advantage for domestic technologies.

This research utilizes survey data from 405 Chinese users to empirically test the moderating effects of institutional trust. The paper is structured as follows: first, we develop the Institutional Prism framework and formulate hypotheses based on Cognitive-Affective Trust Theory. Second, we describe the methodology and present the results of our comparative analysis. Finally, we discuss the theoretical implications of our findings for media sociology and offer practical insights for inclusive AI governance and global technological equity.

## Literature Review and Hypotheses

### *Paradigms of Human-AI Trust from Technical Rationality to Social Agency*

Trust, as a psychological foundation in human–machine collaboration, determines social acceptance, sustained use, and the stability of interactions between users and AI systems (Riegelsberger et al., 2005). In intelligent communication environments, the construction of trust is no longer a purely rational evaluation process but a comprehensive psychological mechanism integrating technological features, social contexts, and cultural affiliation. With the rapid development of computer technology, the study of trust has extended into non-human domains, particularly in the context of human–machine interaction (Y.-K. Lee et al., 2014).

The conceptualization of trust in human-machine communication has traditionally been dominated by two paradigms. The first is the techno-centric paradigm, which can be specifically categorized into four categories of system-level trust antecedents: effectiveness, transparency, explanation, and human oversight (Grimmelikhuijsen, 2022), heavily informed by information systems research, including but not limited to Technology Acceptance Model (TAM) (Davis, 1989). It assumes that people trust AI when it performs well and distrust it when it does not (König et al., 2022). This perspective approaches user trust as a calculative, rational assessment of an AI system's functional attributes, such as accuracy, computational reliability, and transparency. Under a techno-rational view, users may attribute higher cognitive trust to global LLMs because of their perceived scale, sophistication, and performance consistency. Yet cross-national evidence suggests that such capability-based judgments do not fully determine trust (Gillespie et al., 2023).

Conversely, the socio-relational paradigm, rooted in the 'Computers Are Social Actors' (CASA) framework (Nass et al., 1994), which shifts trust formation from performance-based evaluation to socially embedded perception, suggests that as AI systems exhibit increasing autonomy and natural language proficiency, they are evaluated not merely as tools, but as social agents (Nass & Moon, 2000). Scholars emphasize that trust is fostered through anthropomorphic cues, social presence, and cultural alignment (Nass & Moon, 2000; Lee & See, 2004). From this viewpoint, domestic AI models possess an inherent advantage (Sun et al., 2024). By leveraging localized linguistic nuances, culturally specific idioms, and shared societal values, domestic systems may achieve a form of epistemic alignment, transforming AI from a utility into an empathetic partner.

While both paradigms offer critical individual-level insights, current literature largely operates within a structural vacuum, failing to account for the macro-political and institutional forces that govern emerging communication technologies (Flew, 2024). AI systems do not operate independently; they are heavily embedded within national regulatory frameworks, data sovereignty debates, and geopolitical contexts. Recent research increasingly shows that trust in AI is shaped not only by system features but also by trust in governments, firms, and governance institutions (Amir et al., 2025). Therefore, evaluating user trust in domestic versus global AI requires a multidimensional framework that bridges individual psychological assessment with macro-institutional realities.

### *The Cognitive-Affective Trust Theory*

To capture the complexity of AI evaluation, this study adopts the Cognitive-Affective Trust Theory (CATT). Originally proposed by McAllister (1995), CATT posits that trust is not a unidimensional construct but a bipartite psychological structure comprising cognitive and affective dimensions (Benbasat & Wang, 2005).

Cognitive trust refers to a rational belief in the capability, transparency, and reliability of a trustee (Schoorman et al., 2007). In the context of generative AI, users face high 'algorithmic uncertainty' (Kizilcec, 2016). When evaluating cognitive trust, users may face a dilemma: they might recognize the global stability and vast training data of models like ChatGPT (Dwivedi et al., 2021), thereby attributing high computational competence to it. Alternatively, they might perceive domestic models like DeepSeek as possessing higher task-specific reliability and compliance within the local context (Shen et al., 2026). Given the conflicting evidence regarding whether global computational power or local accuracy drives cognitive trust, we propose the following research question:

**RQ1**: Is there a significant difference in Chinese users' cognitive trust between the domestic AI model (DeepSeek) and the global AI model (ChatGPT)?

Affective trust centers on emotional bonds, empathy, and a sense of belonging formed through interaction (McAllister, 1995). In intelligent communication contexts, affective trust is highly contingent upon cultural alignment. Domestic models are specifically optimized for local cognitive patterns, utilizing polite expressions, cultural metaphors, and familiar narrative structures that enhance a user's sense of social presence (Kim et al., 2012). International models may face an affective disadvantage when users perceive weaker cultural alignment. Therefore, we hypothesize:

**H1**: Chinese users will demonstrate higher affective trust in the domestic AI model (DeepSeek) than in the global AI model (ChatGPT).

### ***The Institutional Prism: Refraction Mechanisms of Trust***

While Cognitive-Affective Trust Theory (CATT) distinguishes between cognitive and affective dimensions of trust, it offers limited insight into how these evaluations are formed in broader social and institutional contexts. To capture this process, this study proposes the Institutional Prism framework. Drawing inspiration from Pierre Bourdieu's concept (Bourdieu, 1990; Hilgers & Mangez, 2014) of field refraction, the Institutional Prism refers to the process by which users interpret AI systems through prior beliefs about the legitimacy, competence, and protective capacity of institutions. Rather than assessing AI solely based on technical performance, users may refract their judgments through institutional cues such as governance context, provider origin, and regulatory alignment.

This framework builds on, but also extends, trust transfer research (McKnight et al., 2011). Prior studies have shown that trust in public institutions can transfer to trust in digital services and shape adoption intentions (Horsburgh et al., 2011). More recent work shows that similar mechanisms operate in AI contexts: trust in administrative process, local government, and political leadership can be transferred to trust in AI-enabled government systems (Y.-F. Wang et al., 2024; Kostka et al., 2021). In China, trust in government can mitigate algorithm aversion toward chatbot-based public services (Song et al., 2025). These findings suggest that institutional trust can serve as a contributing factor for technological trust under conditions of uncertainty.

However, trust transfer alone is not sufficient to explain why the same AI system may be interpreted differently across users or contexts. For this reason, the Institutional Prism framework introduces a second mechanism: refraction. In physical optics, refraction occurs when light changes its path as it moves through a medium with a specific refractive index, a property that determines how much the medium can bend incident rays (Born & Wolf, 2013). In this context, refraction refers that institutional trust does not simply increase or decrease trust in AI in a linear way; rather, it shapes which features users attend to, which risks they prioritize, and what meanings they attach to model origin and behavior. Cross-national research shows that trust in AI varies systematically with trust in governments, firms, and scientific institutions (Amir et al., 2025). In this sense, AI systems depend on the social environment in which they are received.

A third element of the framework is the role of institution. For users with high institutional trust, the legitimacy of state governance and domestic regulatory frameworks is successfully transferred to domestic systems (Pande et al., 2025). This institutional endorsement may function as a trust premium, especially when users perceive domestic governance institutions as competent and protective (Y.-F. Wang et al., 2024). Cognitively, it compensates for potential technological gaps, regarding the domestic model as safer, more compliant, and ultimately more reliable. Affectively, the domestic model is refracted as a symbolic extension of national identity and collective security (Kao & Sapp, 2022). In this refractive process, technical attributes are translated into signals of national security, cultural compliance, and institutional reliability. Conversely, for users with low institutional trust, this transfer mechanism may become less effective. In the absence of institutional endorsement, these users are more likely to retreat to a techno-rational evaluation, prioritizing the perceived independence and objectivity of global models. The refraction shifts toward the perceived independence and techno-neutrality of global models like ChatGPT. Therefore, the Institutional Prism explains why the same AI model has perceived different evaluations within the same national population.

To illustrate the mechanisms clearly, we have drawn a model schematic diagram. As illustrated in Figure 1 The institutional prism model of AI trust formation, the Institutional Prism model maps fundamental optical principles onto the cognitive-affective evaluation of AI. In this framework, the Incident Ray represents the objective technical attributes and origins of the AI models such as DeepSeek and ChatGPT. Under conditions of algorithmic opacity, these signals pass through the Institutional Prism, where the level of Institutional Trust serves as the refractive index.

In Figure 1, this refractive process follows two distinct paths: For users with high institutional trust, the prism acts as a dense medium with a high refractive Index (Path 1). Similar to how a prism decomposes white light into a visible spectrum (Newton, 2022), this leads to a Trust Premium for domestic models. The refraction can converge institutional endorsement and regulatory compliance into the cognitive dimension, effectively compensating for perceived technical gaps. Affectively, domestic models can be refracted as symbolic extensions of national identity, resulting in High and Wide trust outcomes across the spectrum.

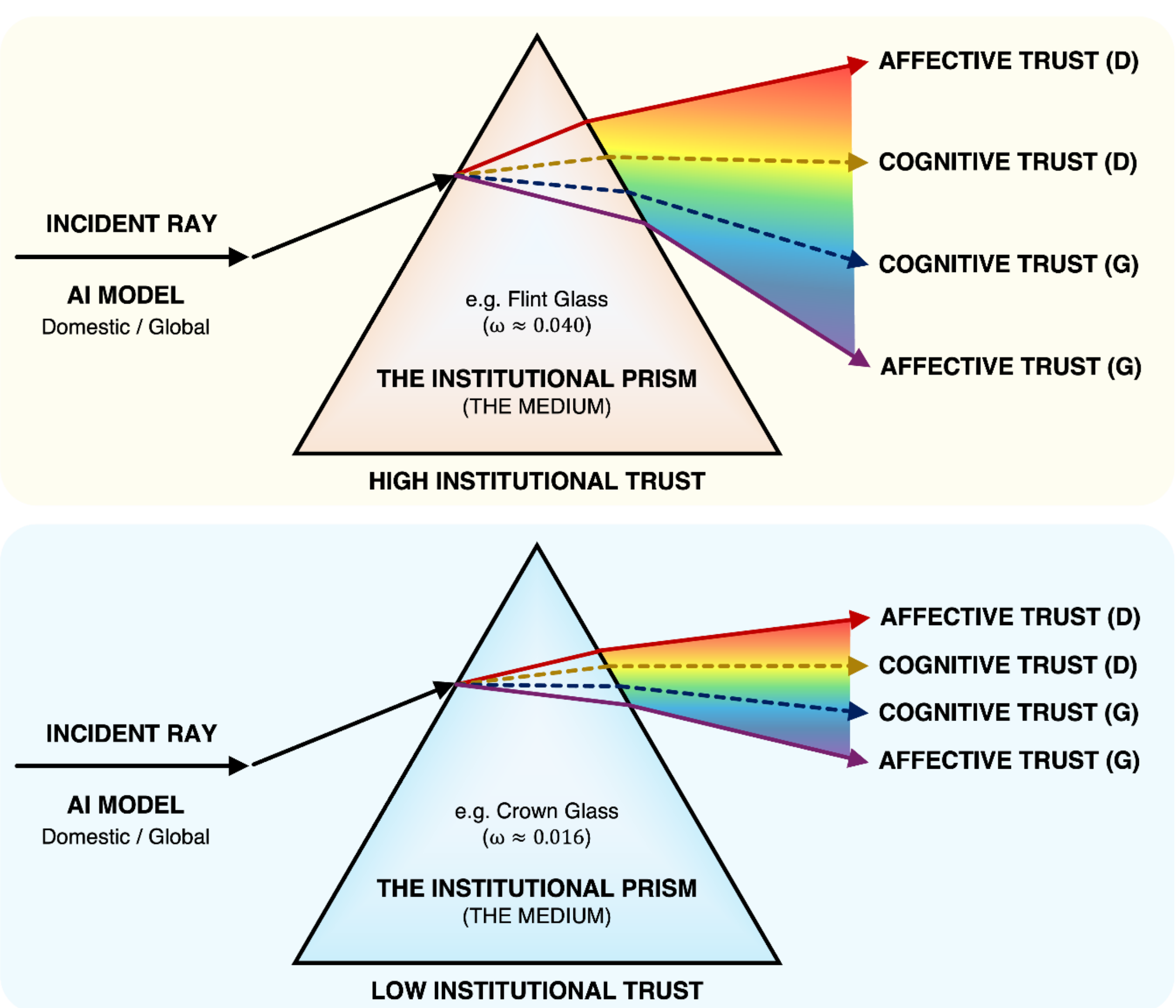

Figure 1 The institutional prism model of AI trust formation

Note: The glass material and dispersion coefficient (ω) are only heuristic metaphors to explain the principle of optical refraction under different institutional trust densities, and do not represent the actual measured mathematical weights.

Conversely, for users with low institutional trust, the prism acts as a weak refractor (Path 2). In the lack of institutional density, the refractive path shifts toward divergent dispersion. Here, institutional cues (such as structural access frictions or domestic regulations) are re-interpreted as constraints on epistemic independence. The evaluation logic diverges from domestic endorsement; the very signals intended to highlight domestic safety are re-interpreted as evidence of global models' epistemic independence. Consequently, the spectrum shifts, where they are Narrow and Focused.

Through this prism, AI systems are institutionally situated artifacts whose final ‘color’ in the trust spectrum is a refracted product of the user's institutional alignment. By moving beyond linear transfer to a refractive logic, this framework explains why the same global technological force results in divergent trust outcomes across different segments of a single national population.

### ***Structural Friction as an Institutional Signal***

In contexts such as China, the Institutional Prism may become especially visible through structural frictions surrounding access to global AI systems, including regulatory restrictions, platform inaccessibility, and other institutional barriers. Rather than treating these frictions as practical obstacles, this study conceptualizes them as institutional cues that make governance boundaries and model origin more salient to users. Prior research suggests that citizens respond not only to technical characteristics of AI, but also to the institutional contexts in which AI is embedded (Schiff et al., 2023).

This matters because structural friction may shape interpretation even before users evaluate model performance. Li's parallel surveys in China and the United States show that respondents trust their own government most and their geopolitical rival least in the domain of global AI governance (Li, 2025). By imposing boundaries on global models, the institutional environment inherently 'others' these foreign technologies, signaling potential risks regarding data sovereignty and informational safety. For high-trust individuals, navigating these frictions may reinforce their alignment with the domestic system and strengthen confidence in AI that appears aligned with local governance structures, deepening their uncertainty toward global models and amplifying their reliance on domestic alternatives (Shen et al., 2026). For low-trust individuals, however, these same structural frictions may create a different valuative pathway, where restriction may paradoxically enhance the symbolic appeal of global models because of its perceived distance from domestic regulation.

Therefore, structural friction does not directly determine trust, but it intensifies the influence of institutional factors on the interpretation of model types. Accordingly, this study does not treat structural friction as an independent variable in the empirical model. Instead, it is theorized as a contextual mechanism within the Institutional Prism: a condition that sharpens how institutional trust moderates users' cognitive and affective evaluations of domestic and global AI systems.

Against this backdrop, institutional trust serves as a measurable mechanism that captures how such contextual conditions translate into differences in users' cognitive and affective evaluations. Accordingly, this study focuses on the moderating role of institutional trust in shaping trust toward domestic and global AI systems. We propose the following hypotheses regarding its moderating effects:

**H2**: Institutional trust moderates the relationship between AI model type and cognitive trust. Specifically, the gap in cognitive trust between domestic and global models diminishes or reverses as institutional trust increases.

**H3**: Institutional trust moderates the relationship between AI model type and affective trust. Specifically, higher institutional trust amplifies the positive predictive effect of DeepSeek on affective trust relative to ChatGPT.

The proposed theoretical model is illustrated in Figure 2.

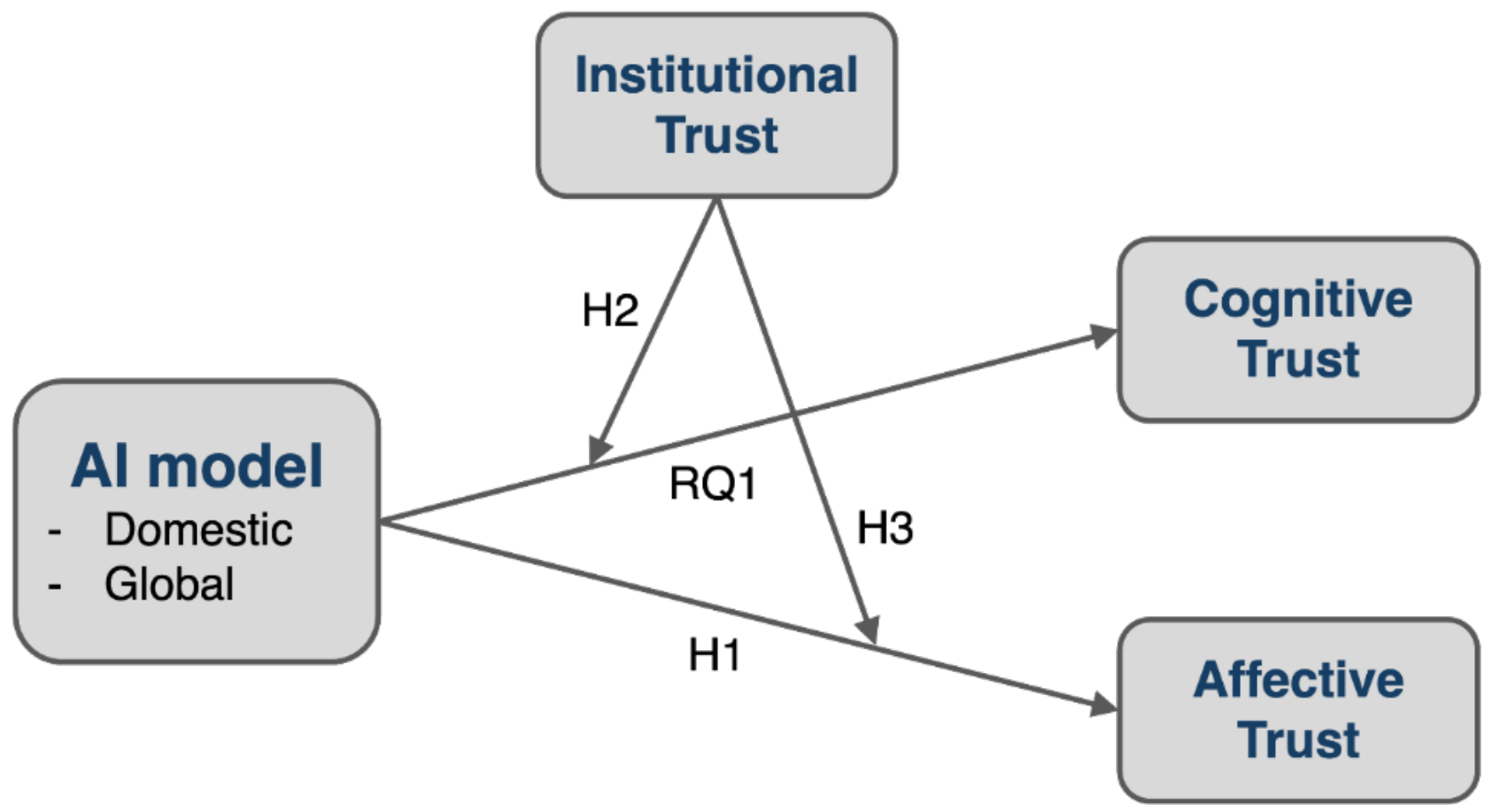

Figure 2 The theoretical model diagram of AI trust formation and institutional moderation.

## Methodology

### *Sample and Procedure*

This study was conducted from February to April 2025, with data collected via the Credamo platform, a leading professional online survey platform in China, which handled questionnaire design, distribution, and collection. A total of 476 questionnaires were collected. After removing invalid responses that failed the attention check, 405 valid samples remained. Since the study includes 13 observation variables, the sample size is adequate for regression-based analyses.

This study focuses on users who are able or consciously evaluate between the two models. Participants had to meet the following criteria: (1) be aged 18 or older, (2) reside in mainland China, and (3) have used at least one domestic AI product (e.g., DeepSeek) and one international AI product (e.g., ChatGPT) in the past three months, allowing for valid trust comparisons across models.

Although skewed toward younger students, this group represents a technical open demographic in AI adoption and intensive usage contexts, showing higher usage frequency, greater technology readiness, and heightened sensitivity to emerging AI, ensuring the sample's validity within AI environment. Despite regional restrictions, global model's usage remains prevalent among highly educated Chinese cohorts via VPNs and mirror APIs. To guarantee data integrity, we enforced a rigorous screening protocol, including attention checks, IP filtering, and response duration monitoring. The study was approved by Wuhan University ethics committee with strict adherence to ethical research standards maintained.

### *Variable Measurement*

The primary variables in this study were AI model type (DeepSeek vs. ChatGPT), cognitive trust, affective trust, and institutional trust, all measured on a 5-point Likert scale from 1 (strongly disagree) to 5 (strongly agree).

**Cognitive Trust**: This variable was assessed using an adapted version of McAllister's (1995) cognitive trust subscale, modified for AI system contexts. Sample items included: 'DeepSeek demonstrates high professionalism and reliability, effectively completing tasks,' 'Based on DeepSeek's past performance, I believe it is capable of handling complex tasks,' and 'In actual use, DeepSeek operates accurately and reliably, with minimal technical issues.' The scale showed good reliability (Cronbach's $\alpha = 0.789$).

**Affective Trust**: Affective trust was assessed using an adapted version of McAllister's (1995) affective trust subscale. Example items included: 'I am willing to share my work progress, emotional state, and creative ideas with DeepSeek,' 'When I encounter difficulties, I trust DeepSeek to understand my needs and provide support,' and 'When submitting feedback or encountering issues, DeepSeek offers empathetic and specific responses.' This scale also showed good reliability (Cronbach's $\alpha = 0.728$).

**Institutional Trust**: This variable reflects the public's trust in government and its actions, specifically concerning AI development. The scale, adapted from Sun et al. (2007), assessed public trust in government institutions and AI-related policies. Example items included: 'The government can mobilize resources to address AI development challenges,' 'AI governance policies are genuinely implemented, not merely for show,' and 'Government officials responsible for AI governance are trustworthy.' This scale demonstrated good reliability (Cronbach's $\alpha = 0.838$).

**Control Variables**: Demographic variables, including age, gender, education level, and occupation, were controlled in regression and moderation models to account for potential confounding effects.

### *Analytical Methods*

This study employed SPSS software for primary data analysis and hypothesis testing. To ensure methodological rigor and fully address potential sampling biases, all key inferential results were systematically cross-checked and validated in Python. The supplementary validation pipeline included Linear Mixed-Effects Modeling (LMM), student versus non-student heterogeneity analysis, post-hoc power analysis, Bonferroni adjustments, and non-parametric robust testing.

### *Common Method Bias*

As this was a cross-sectional study, data were collected simultaneously, which may introduce common method bias (CMB) and potentially affect the accuracy of the results. To minimize this risk, several measures were adopted during the study's design and implementation phases. First, anonymous questionnaires were administered to ensure respondent privacy and reduce social desirability bias. Second, the questionnaire was designed to avoid vague or double-barreled items, promoting clarity and precision. Different measurement scales were employed for different latent variables to minimize systematic error arising from single-source measurement tools. Additionally, during data analysis, Harman's Single Factor Test was performed in SPSS to assess potential common method bias (Podsakoff et al., 2003). The results indicated that the variance explained by the first unrotated factor was 32%, below the 40% threshold, suggesting that common method bias was limited.

## Results and Findings

### ***Descriptive Statistics and Correlation Analysis of Key Variables***

Preliminary examination of the correlation coefficients between variables showed that all correlations were below 0.7, effectively mitigating potential multicollinearity issues. Descriptive statistics, including means and standard deviations, are presented in Table 1 Pearson Correlation Matrix of Key Variables. To further examine relationships among the variables, the Pearson correlation matrix for the key variables is presented. The results indicate significant positive correlations among AI model type, affective trust, and cognitive trust, with institutional trust demonstrating strong associations with both affective and cognitive trust.

Table 1 Pearson Correlation Matrix of Key Variables

| **Variable** | **Mean** | **SD** | **AI Type** | **Affective Trust** | **Cognitive Trust** | **Institutional Trust** |
|---|---|---|---|---|---|---|
| **AI Model** | 0.500 | 0.500 | 1 | - | - | - |
| **Affective Trust** | 3.310 | 0.693 | 0.188** | 1 | - | - |
| **Cognitive Trust** | 3.541 | 0.623 | 0.065 | 0.568** | 1 | - |
| **Institutional Trust** | 3.672 | 0.769 | 0.000 | 0.222** | 0.276** | 1 |

**$p < 0.01$

### ***Paired-Sample T-Test***

To explore the differences in users' trust levels under different AI models (DeepSeek and ChatGPT), this study conducted a paired-sample t-test on the scores of the same participants across two dimensions: cognitive trust and affective trust.

In terms of cognitive trust, the mean cognitive trust score of DeepSeek was 3.58 (SD = 0.59), while that of ChatGPT was 3.51 (SD = 0.66). The paired-sample t-test results indicated a significant difference in cognitive trust between the two models ($t = 2.474$, $p < 0.05$), showing that Chinese users had higher cognitive trust in DeepSeek than in ChatGPT.

While in terms of affective trust, the mean affective trust score of DeepSeek was 3.44 (SD = 0.73), while that of ChatGPT was 3.18 (SD = 0.63). The paired-sample t-test also revealed a significant difference ($t = 6.697$, $p < 0.001$), further showing that Chinese users had a higher level of affective trust in DeepSeek than in ChatGPT.

Consequently, users exhibited higher trust in DeepSeek across both cognitive and affective dimensions, with a stronger preference for affective trust. The detailed paired-comparison results are summarized in Table 2 Paired-sample t-test results for DeepSeek vs. ChatGPT.

Table 2 Paired-sample t-test results for DeepSeek vs. ChatGPT

| Measure | DeepSeek M (SD) | ChatGPT M (SD) | Δ | t | p | Cohen's d |
|---|---|---|---|---|---|---|
| **Cognitive Trust** | 3.582 (0.586) | 3.501 (0.655) | 0.081 | 2.474 | 0.014 | 0.130 |
| **Affective Trust** | 3.440 (0.727) | 3.179 (0.631) | 0.261 | 6.697 | <0.001 | 0.383 |

***Moderating Effect of Institutional Trust***

To examine whether institutional trust moderates the relationship between AI type and user trust, the PROCESS macro (Model 1) developed by Hayes (2017) was employed. AI type (DeepSeek vs. ChatGPT) served as the independent variable, institutional trust as the moderator, and cognitive trust (H2) and affective trust (H3) as the dependent variables, to examine the significance of the interaction effect. A total of 5000 bootstrap samples were generated, with a 95% confidence interval. Additionally, simple slope analysis was conducted on significant interaction terms to illustrate how AI type affects user trust at different levels of institutional trust.

Results indicated that the interaction term (AI type × institutional trust) significantly moderated cognitive trust ($p < 0.05$). Specifically, at a low level of institutional trust (M − 1 SD), the difference in cognitive trust between DeepSeek and ChatGPT was minimal (M_DeepSeek = 3.36, M_ChatGPT = 3.38), whereas at a high level of institutional trust (M + 1 SD), the difference increased (M_DeepSeek = 3.81, M_ChatGPT = 3.62). These results suggest institutional trust positively moderates the relationship between AI type and cognitive trust: as institutional trust increases, DeepSeek's cognitive trust advantage becomes more pronounced. This finding supports H2.

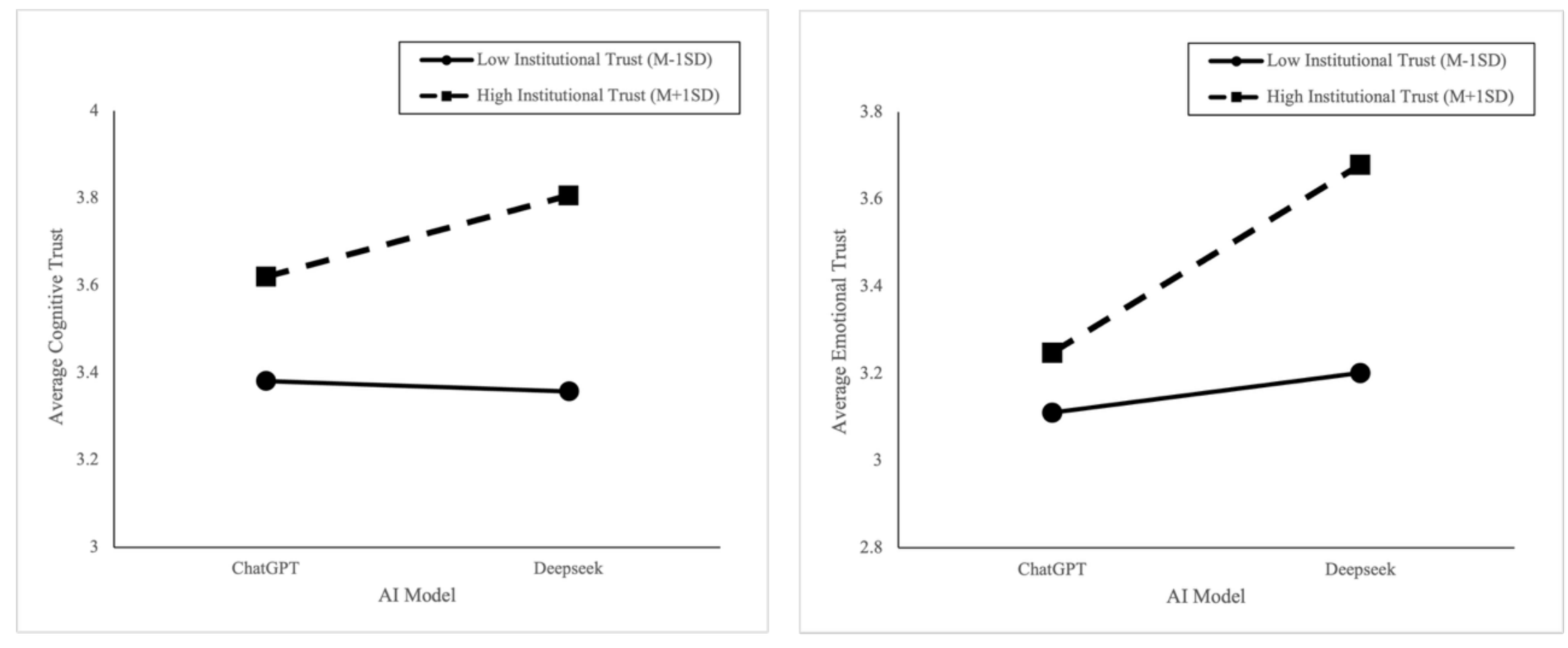

(a) The moderating effect of cognitive trust (b) The moderating effect of affective trust

Figure 3 The schematic diagram of the moderating effect

We also analyzed the moderating role of institutional trust between AI type and affective trust. The interaction term between AI type and institutional trust also significantly moderates affective trust ($p < 0.001$). Specifically, at a low level of institutional trust, the difference in affective trust between DeepSeek and ChatGPT was minimal (M_DeepSeek = 3.20, M_ChatGPT = 3.11), whereas at a high level of institutional trust, DeepSeek's affective trust score increased substantially (M_DeepSeek = 3.68, M_ChatGPT = 3.25). This result indicates that institutional trust positively moderates the relationship between AI type and affective trust. In contexts of high institutional trust, DeepSeek is more likely to elicit greater affective trust and emotional endorsement from users. This finding supports H3.

As illustrated in Figure 3 The schematic diagram of the moderating effect, the findings indicate that institutional trust plays a critical moderating role between AI type and user trust.

### *Robustness Checks and Supplementary Validations*

To guarantee that the primary empirical conclusions were robust against potential sampling anomalies, demographic confounders, or specific statistical assumptions, a multi-stage validation framework was implemented in Python.

As shown in Table 3 Heterogeneity test: student versus non-student comparison,we tested whether the main AI-type effects differed between student and non-student respondents. The results showed no significant differences in cognitive trust, affective trust, or institutional trust between the two groups, while the AI-origin effect remained visible in both subsamples. This indicates that the observed DeepSeek advantage is not confined to a single demographic cluster and therefore supports the external validity of the main findings.

Table 3 Heterogeneity test: student versus non-student comparison

| Variable | Student M (SD) | Non-student M (SD) | t | p | Cohen’s d |
|---|---|---|---|---|---|
| **Cognitive Trust** | 3.521 (0.605) | 3.578 (0.654) | -1.252 | 0.211 | -0.092 |
| **Affective Trust** | 3.300 (0.692) | 3.328 (0.697) | -0.552 | 0.581 | -0.041 |
| **Institutional Trust** | 3.665 (0.742) | 3.685 (0.818) | -0.352 | 0.725 | -0.026 |
| **Age** | 19.275 (1.622) | 37.545 (10.533) | -38.890 | <0.001 | -2.859 |

At the same time, the large age gap between students and non-students confirms that the heterogeneity test is also an effective control for potential confounding. In other words, if the pattern had been driven primarily by age or life-stage composition, the student/non-student comparison would have produced substantial divergence; instead, the effect pattern remains stable, suggesting that the main conclusions are not an artifact of sample composition.

Table 4 Mixed effects model results in long-format data

| Dependent variable | Fixed effect | β | z/t | p | Interpretation |
|---|---|---|---|---|---|
| **Cognitive Trust** | Model Type | 0.041 | 2.474 | 0.013 | Significant |
| **Cognitive Trust** | Institutional Trust | 0.172 | 6.918 | <0.001 | Significant |
| **Cognitive Trust** | Age | 0.017 | 0.400 | 0.689 | Not significant |
| **Cognitive Trust** | Gender | -0.063 | -1.227 | 0.220 | Not significant |
| **Cognitive Trust** | Student | 0.020 | 0.231 | 0.818 | Not significant |
| **Affective Trust** | Model Type | 0.131 | 6.697 | <0.001 | Significant |
| **Affective Trust** | Institutional Trust | 0.154 | 5.750 | <0.001 | Significant |
| **Affective Trust** | Age | 0.014 | 0.300 | 0.764 | Not significant |
| **Affective Trust** | Gender | 0.054 | 0.973 | 0.331 | Not significant |

| Dependent variable | Fixed effect | β | z/t | p | Interpretation |
|---|---|---|---|---|---|
| **Affective Trust** | Student | 0.005 | 0.049 | 0.961 | Not significant |

To further control confounding and account for the repeated-measures structure of the data, the wide-format dataset was reshaped into long format and estimated with a linear mixed effects model (Table 4 Mixed effects model results in long-format data). In this specification, each participant contributed two observations, with subject included as a random intercept and model type, institutional trust, age, gender, and student status entered as fixed effects. This approach preserves within-person dependence while simultaneously testing the robustness of the main effects under covariate control.

The model confirmed the same substantive pattern as the earlier analyses. For cognitive trust, model type remained significant ($\beta = 0.041$, $p = 0.013$), and institutional trust was a strong positive predictor ($\beta = 0.172$, $p < 0.001$), whereas age, gender, and student status were not significant. For affective trust, the model-type effect was even stronger ($\beta = 0.131$, $p < 0.001$), institutional trust again remained significant ($\beta = 0.154$, $p < 0.001$), and the control variables were still negligible. Thus, the mixed model provides a stronger identification strategy and reinforces the claim that the observed AI-origin effect is not explained by simple sample composition or demographic confounding.

Meanwhile, post-hoc power analysis showed that the affective-trust effect was highly powered (100.0%), whereas the smaller cognitive-trust effect remained more modest but still acceptable (74.3%). This suggests that the refractive effect of institutional prism is mainly driven by emotional trust, while differences in technical evaluation at the cognitive level are relatively marginal.

## Discussion

This study employed the Cognitive-Affective Trust Theory and the Institutional Prism framework to examine the trust mechanisms underlying domestic and global large language models. The empirical results suggest that trust is a socially embedded process in which institutional legitimacy shapes both perceived competence and affective resonance.

### *The Asymmetry of Trust: Why Affective Resonance Outpaces Cognitive Evaluation*

The findings show that domestic AI models command a distinct trust premium over global models, a divergence that is particularly pronounced in the affective dimension. Despite DeepSeek's relatively brief development trajectory, its structural optimization for the Chinese social landscape has significantly elevated users' evaluations of its cognitive capacity and operational reliability. Another pivotal contribution of this study is revealing the asymmetric moderating effect of institutional trust on the two dimensions of Cognitive-Affective Trust Theory. Specifically, the data indicates that institutional trust exerts a profoundly stronger influence on affective trust than on cognitive trust. How can we explain this divergence?

This asymmetry is deeply rooted in the distinct psychological thresholds required to form cognitive versus affective bonds with artificial intelligence. Recent comprehensive reviews on human-AI trust emphasize that cognitive trust is predominantly driven by observable computational reliability and functional performance (Glikson & Woolley, 2020). In the contemporary digital landscape, the massive technological prowess and systemic stability of global models like ChatGPT are widely acknowledged epistemic realities. Consequently, even users with high domestic institutional trust cannot entirely dismiss the functional competence of global algorithms, leading to a relatively narrower gap in cognitive trust evaluations.

Conversely, affective trust implies vulnerability, reliance, and a perception of social values (Glikson & Woolley, 2020). When users lack the algorithmic literacy to audit AI 'black boxes', they increasingly rely on macro-institutional proxy signals to determine whether an AI system is fundamentally safe to emotionally rely upon (Araujo et al., 2020). While generative AI is globally developed, its operationalization involves localized data annotation and manual alignment (Fu et al., 2025). For users with high institutional trust, these regulatory behaviors function as powerful signals of algorithmic sovereignty and identity validation. Domestic models are thus perceived as 'in-group social actors' that respect local norms. This translates macro-level state legitimacy into affective premium, explaining why emotional attachment exhibits significantly higher sensitivity to institutional cues than cognitive evaluation.

### *The Ideological Buffer: Recontextualizing Algorithmic Tolerance*

A further theoretical implication of our findings lies in how users perceive algorithmic constraints. Mainstream research frequently highlights the phenomenon of algorithm aversion, suggesting that users lose trust in AI systems when algorithms exhibit rigid boundaries, constraints, or filtering mechanisms (Dietvorst et al., 2015; Jussupow et al., 2020). However, the robust trust premium awarded to domestic models by high-institutional-trust users in our study challenges this universalist assumption, revealing a phenomenon of ideological buffering.

Given the regulatory boundaries and content filtering inherent in domestic large language models, one might conventionally expect a depletion in user trust due to perceived respondent limitations. Yet, our empirical data demonstrates the exact opposite: high institutional trust significantly elevates both cognitive and affective endorsement. This suggests that macro-level institutional alignment acts as a powerful cognitive buffer. For them, the systemic constraints of domestic AI are not penalized as incompetence or information deprivation. Instead, users actively rationalize these boundaries as manifestations of normative safety and cultural alignment.

In this paradigm, institutional trust generates a profound tolerance. When the AI's boundaries align with the user's macro-institutional worldview, functional limitations are recontextualized as protective affordances rather than technical flaws. Therefore, in highly institutionalized digital ecosystems, a model's credibility is not solely derived from its unconstrained generative capabilities, but from its capacity to operate safely within the ideological contours expected by its users.

### *The Liability of Foreignness: Divergence and information Balkanization*

Finally, moving beyond individual psychological processing, our findings suggest profound macro-level implications for the global governance of generative AI. The moderating effect of institutional trust indicates that the development of AI may not necessarily align with the vision of a borderless, universal knowledge infrastructure. Because trust is deeply intertwined with perceived sovereign legitimacy, global models face an inherent liability of foreignness (Zaheer, 1995; O'Hara & Hall, 2021) when crossing geopolitical boundaries.

No matter how computationally superior a global system may be, its perceived lack of local institutional embeddedness may limit its affective trust capacity among users who strongly align with their domestic regimes. This suggests the potential emergence of information balkanization in the age of AI-a condition where the digital ecosystem is structurally partitioned into non-communicating knowledge islands. The Institutional Prism does not merely influence user preference; it likely facilitates a structural partition of the digital ecosystem into divergent realities. Users with differing institutional orientations may increasingly favor models that reinforce their existing worldviews, leading to a tendency toward algorithmic self-segregation. Consequently, rather than fostering global cognitive convergence, the proliferation of large language models, when refracted through national institutional prisms, may deepen the fragmentation of the global knowledge ecosystem and encourage the formation of localized epistemic enclosures.

### ***Theoretical Contributions***

This study advances the theoretical contours of human-machine communication by dismantling the prevailing dyadic assumption. While existing research has largely conceptualized user trust as a direct combination with interface design and computational capability (Hancock et al., 2011; Glikson & Woolley, 2020), our findings demonstrate that this interaction is profoundly mediated by macro-political environments. By introducing the concept of institutional refraction into the AI trust paradigm (Hilgers & Mangez, 2014; Couldry, 2003), we theorize that structural environments function as critical epistemic lenses rather than mere background contexts. This framework offers a theoretical mechanism to explain why globally computational architectures, such as generative large language models, are locally reinterpreted and legitimized through divergent, culturally situated credibility assessments.

Building upon this, this research extends the Cognitive-Affective Trust Theory (CATT) by conceptualizing a macro-to-micro trust transfer mechanism. Previous research on trust, heavily anchored in interpersonal and organizational psychology (McAllister, 1995), has isolated the rational and emotional dimensions of trust from broader sociopolitical governance beliefs. Our empirical evidence structurally challenges this isolation. We illustrate that cognitive and affective trust in artificial intelligence do not form in a vacuum; instead, macro-level institutional trust acts as a signal. By operating institutional trust as an upstream factor of algorithmic credibility, this study reveals the dynamic process through which macro-level state legitimacy is translated into micro-level user endorsement.

Consequently, this research prompts a new evaluation within global communication discourse regarding the nature of geopolitical technological barriers. It reframes structural access friction such as regulatory boundaries, not as mere technological impediments or communicative deficits, but as potent epistemic signals for local (Bucher, 2017). Where existing literature frequently treats these frictions as systemic noise hindering global AI adoption, our comparative analysis between domestic and global models demonstrates that it has differential impacts on people with different levels of institutional trust. As evidenced by the persistence of these refracted trust patterns, AI models are encountered not as culturally neutral computational engines, but as institutionally situated communicative artifacts (Flew, 2024). Ultimately, this study demonstrates that institutional boundaries fundamental components of an AI system's sociotechnical nature and its capacity to command public trust.

### *Practical Implications*

From a governance perspective, the findings highlight that policymakers must recognize trust not merely as a byproduct of technological advancement, but as a crucial psychosocial resource that requires strategic management. In societies characterized by high institutional trust, governments can leverage their symbolic legitimacy to foster early adoption of domestic innovations. However, policymakers need to be vigilant against the risk of cultivating closed epistemic ecosystems. To prevent cognitive echo chambers, regulatory frameworks should balance the promotion of domestic algorithmic security with the facilitation of cross-border interoperability and diverse knowledge integration, understanding that trust management needs governance transparency, not only technical optimization.

For AI developers and industry practitioners, this study underscores that achieving algorithmic competitiveness in localized markets requires far more than mere linguistic translation or interface localization. Developers should prioritize institutional alignment as a component of system design. Building affective trust necessitates the integration of culturally congruent interaction logic and transparent compliance with local ethical norms. Furthermore, developers of global AI models need to adopt participatory design frameworks that actively address the skepticism of local users, establishing culturally responsive and institutionally sensitive technological solutions.

### ***Limitations and Future Research***

Several limitations warrant consideration for future research. The cross-sectional design restricts the ability to track the temporal evolution of trust formation. Given that generative AI is a rapidly advancing technology and geopolitical environments are highly dynamic, longitudinal studies are essential to capture how shifting institutional narratives and policy changes iteratively reshape user trust over time.

Additionally, the sample predominantly comprises younger populations who are typically early adopters of AI technologies. While this demographic offers critical insights into emerging digital behaviors, their sociopolitical experiences and interaction frequencies with institutional systems may differ significantly from those of older generations. Future research should strive for broader demographic representation to test the robustness of the Institutional Prism model across various age cohorts and socioeconomic strata.

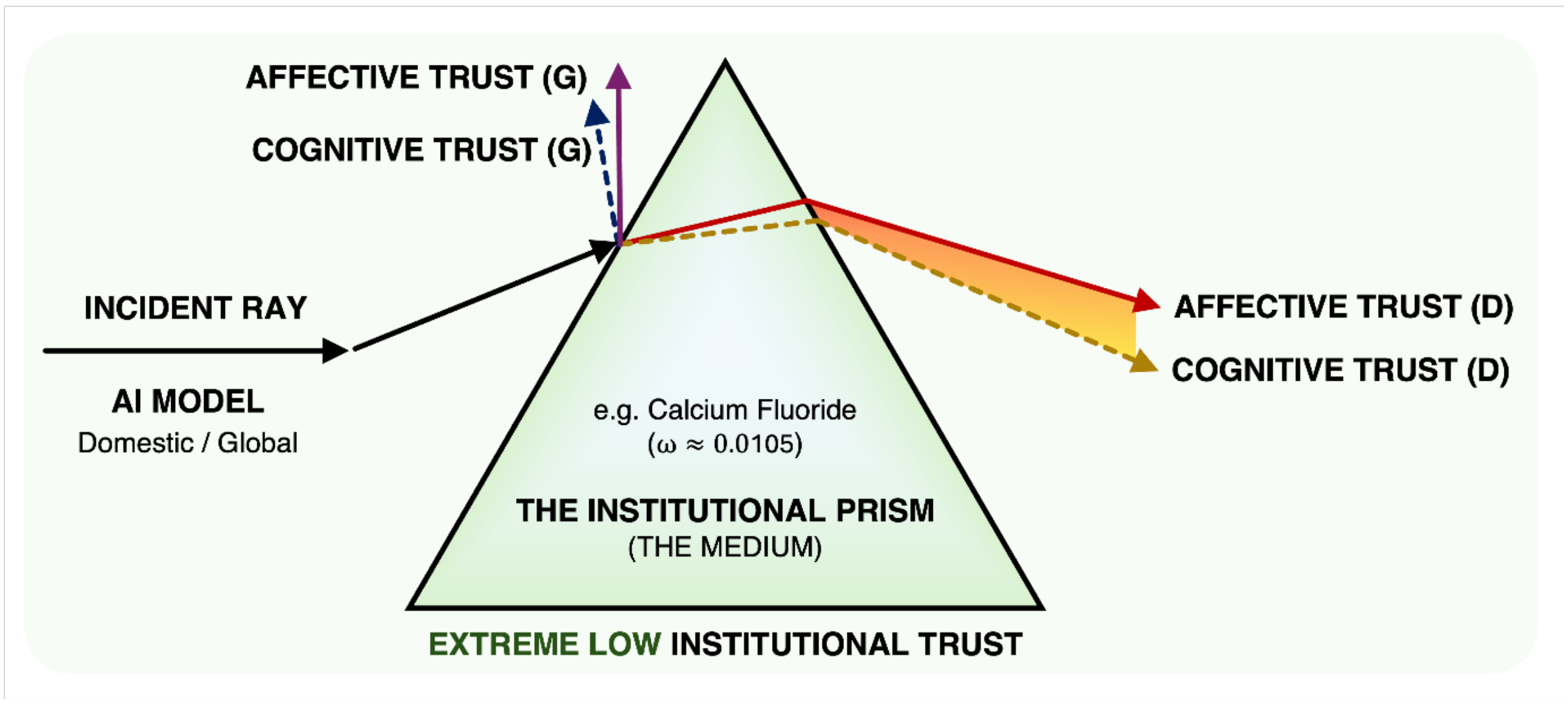

Figure 4 Total reflection of the institutional prism model

Moreover, while this study identifies the divergent refraction path for low-trust users, future inquiries could explore the model's extreme boundary conditions. In optical physics, total internal reflection occurs when light traveling through a dense medium strikes a boundary at an angle exceeding the critical angle, causing the ray to reflect entirely back into the original medium without refraction (Born & Wolf, 2013). In a socio-technical context, this represents a critical threshold of distrust. When institutional trust falls below this tipping point, the institutional prism may no longer refract perceptions but instead trigger total cognitive rebound. Users completely deflect external algorithmic inputs regardless of their objective technological competence. Exploring whether such a critical angle exists across different political cultures would deepen our understanding of how polarized institutional fields govern the flow of trust in global communication.

Finally, as this study is situated within the Chinese societal context, the generalizability of the institutional refraction mechanism requires further empirical validation. Comparative studies between high- and low-trust institutional settings would help clarify how diverse governance paradigms moderate the relationship between AI origin and user trust, ultimately revealing the broader structural forces governing global communication inequalities.

# Appendix

## A Complete Questionnaire

**Dear Sir/Madam,**

Hello,

We are conducting an online survey to explore users' experiences with AI chat tools. Your responses will help researchers better understand user preferences across different AI systems and provide valuable insights for the optimization of AI technology.

Please answer each question honestly based on your actual experiences and true feelings. The questionnaire is expected to take approximately 5–10 minutes to complete. Feel free to finish it at your convenience.

This survey is completely anonymous, and your personal information will be kept strictly confidential.

There are no right or wrong answers. All responses will be used solely for academic research purposes. We greatly value your opinions.

**(Section 1)** Please fill in the following information based on your actual situation.

**Q1** What is your age? [Open-ended]

**Q2** What is your gender? [Single Choice]

- Male
- Female

**Q3** What is your occupation type? [Single Choice]

- Student
- Internet/IT Industry
- Education/Research
- Finance/Law
- Entertainment/Arts
- Other

**Q4** What is your highest level of education? [Single Choice]

- Primary school or below
- Junior high school
- Senior high school (including technical secondary school)
- Junior college (Associate degree)
- Bachelor's degree
- Master's degree
- Doctoral degree

**Q5** What are the main purposes for which you use AI tools most frequently? [Multiple Choice]

- Searching for information/news
- Assisting with writing/translation
- Coding/programming
- Entertainment (chatting, image generation, etc.)

- Other

**Q6** When using AI tools, which of the following aspects do you care about most? [Ranking]

Please rank them in order of importance.

- Response speed
- Professionalism and accuracy
- Data security and privacy
- Interaction experience
- Other (please specify: ____)

**Q7** How often do you use ChatGPT? [Single Choice]

- Never
- Rarely
- Sometimes
- Often
- Always

**Q8** How often do you use DeepSeek? [Single Choice]

- Never
- Rarely
- Sometimes
- Often

- Always

**Q9** On average, how many hours per week do you use AI tools (such as ChatGPT, DeepSeek)? [Single Choice]

- Almost never
- 1–5 hours
- 6–10 hours
- More than 10 hours

**Q10** The government can fully mobilize and coordinate all forces to address the challenges brought by AI development. [Scale]

(1 = Strongly Disagree, 5 = Strongly Agree)

○ 1 ○ 2 ○ 3 ○ 4 ○ 5

**Q11** Policies related to the governance of AI development are earnestly implemented rather than remaining superficial. [Scale]

(1 = Strongly Disagree, 5 = Strongly Agree)

○ 1 ○ 2 ○ 3 ○ 4 ○ 5

**Q12** In addressing the challenges of AI development, government officials are trustworthy. [Scale]

(1 = Strongly Disagree, 5 = Strongly Agree)

○ 1 ○ 2 ○ 3 ○ 4 ○ 5

**Q13** The policies of the Party and the government genuinely care about the interests of ordinary citizens. [Scale]

(1 = Strongly Disagree, 5 = Strongly Agree)

○ 1 ○ 2 ○ 3 ○ 4 ○ 5

**Q14** The model Party members and cadres established by the Party and the government enjoy high prestige. [Scale]

(1 = Strongly Disagree, 5 = Strongly Agree)

○ 1 ○ 2 ○ 3 ○ 4 ○ 5

**Q15** The Party and the government enjoy high prestige among ordinary citizens. [Scale]

(1 = Strongly Disagree, 5 = Strongly Agree)

○ 1 ○ 2 ○ 3 ○ 4 ○ 5

**Q16** The Party and the government have the willingness to uphold justice for ordinary citizens. [Scale]

(1 = Strongly Disagree, 5 = Strongly Agree)

○ 1 ○ 2 ○ 3 ○ 4 ○ 5

**Q17** The Party and the government have the ability to uphold justice for ordinary citizens. [Scale]

(1 = Strongly Disagree, 5 = Strongly Agree)

○ 1 ○ 2 ○ 3 ○ 4 ○ 5

**(Section 2) DeepSeek Section**

**Q0** DeepSeek is an AI assistant developed by the DeepSeek AI team. It specializes in Chinese and multilingual text processing, with strong capabilities in information comprehension and generation. It can be used for intelligent Q&A, content creation, code assistance, text summarization, and other tasks. It is suitable for writing, learning, research, and programming needs. DeepSeek provides online access, allowing users to interact through the corresponding platform to obtain more efficient knowledge support and task processing capabilities.

**Q1** I am willing to share my work progress, emotional state, and innovative ideas with DeepSeek. [Scale]

(1 = Strongly Disagree, 5 = Strongly Agree)

○ 1 ○ 2 ○ 3 ○ 4 ○ 5

**Q2** When I encounter difficulties, I am willing to report problems to DeepSeek and believe it can understand my needs and provide support. [Scale]

(1 = Strongly Disagree, 5 = Strongly Agree)

○ 1 ○ 2 ○ 3 ○ 4 ○ 5

**Q3** If DeepSeek temporarily suspends service due to technical upgrades, I would feel inconvenienced and upset. [Scale]

(1 = Strongly Disagree, 5 = Strongly Agree)

○ 1 ○ 2 ○ 3 ○ 4 ○ 5

**Q4** When I submit feedback or encounter problems with DeepSeek, its responses are specific and empathetic (e.g., “We noticed you encountered difficulties while using the XX feature”). [Scale]

(1 = Strongly Disagree, 5 = Strongly Agree)

○ 1 ○ 2 ○ 3 ○ 4 ○ 5

**Q5** If DeepSeek temporarily suspends service due to technical upgrades, I would **not** feel inconvenienced or upset. [Scale]

(1 = Strongly Disagree, 5 = Strongly Agree)

○ 1 ○ 2 ○ 3 ○ 4 ○ 5

**Q6** Before I started using DeepSeek, I already had a certain level of trust in its professionalism and reliability. [Scale]

(1 = Strongly Disagree, 5 = Strongly Agree)

○ 1 ○ 2 ○ 3 ○ 4 ○ 5

**Q7** DeepSeek has high professionalism and reliability and can effectively complete tasks. [Scale]

(1 = Strongly Disagree, 5 = Strongly Agree)

○ 1 ○ 2 ○ 3 ○ 4 ○ 5

**Q8** Based on DeepSeek's past performance, I believe it is good at handling complex tasks. [Scale]

(1 = Strongly Disagree, 5 = Strongly Agree)

○ 1 ○ 2 ○ 3 ○ 4 ○ 5

**Q9** In actual use, DeepSeek operates accurately and reliably, with few technical issues disrupting my workflow. [Scale]

(1 = Strongly Disagree, 5 = Strongly Agree)

○ 1 ○ 2 ○ 3 ○ 4 ○ 5

**Q10** My friends and colleagues generally believe that DeepSeek has high professionalism and credibility. [Scale]

(1 = Strongly Disagree, 5 = Strongly Agree)

○ 1 ○ 2 ○ 3 ○ 4 ○ 5

**Q11** As I gain a deeper understanding of AI technology principles and background, I pay more attention to DeepSeek's performance. [Scale]

(1 = Strongly Disagree, 5 = Strongly Agree)

○ 1 ○ 2 ○ 3 ○ 4 ○ 5

**(Section 3) ChatGPT Section**

**Q0** ChatGPT is an AI conversational assistant developed by OpenAI. It can handle various text-based communication tasks and is suitable for daily conversations, knowledge Q&A, writing assistance, code generation, language translation, and many other scenarios. Users can access ChatGPT through web pages or mobile applications and interact with it in a conversational manner to obtain information, write articles, refine ideas, or solve problems. ChatGPT has wide applications in content creation, office productivity, and learning support, helping users enhance their productivity and creativity.

**Q1** I am willing to share my work progress, emotional state, and innovative ideas with ChatGPT. [Scale]

(1 = Strongly Disagree, 5 = Strongly Agree)

○ 1 ○ 2 ○ 3 ○ 4 ○ 5

**Q2** When I encounter difficulties, I am willing to report problems to ChatGPT and believe it can understand my needs and provide support. [Scale]

(1 = Strongly Disagree, 5 = Strongly Agree)

○ 1 ○ 2 ○ 3 ○ 4 ○ 5

**Q3** If ChatGPT temporarily suspends service due to technical upgrades, I would feel inconvenienced and upset. [Scale]

(1 = Strongly Disagree, 5 = Strongly Agree)

○ 1 ○ 2 ○ 3 ○ 4 ○ 5

**Q4** When I submit feedback or encounter problems with ChatGPT, its responses are specific and empathetic (e.g., “We noticed you encountered difficulties while using the XX feature”). [Scale]

(1 = Strongly Disagree, 5 = Strongly Agree)

○ 1 ○ 2 ○ 3 ○ 4 ○ 5

**Q5** If ChatGPT temporarily suspends service due to technical upgrades, I would **not** feel inconvenienced or upset. [Scale]

(1 = Strongly Disagree, 5 = Strongly Agree)

○ 1 ○ 2 ○ 3 ○ 4 ○ 5

**Q6** Before I started using ChatGPT, I already had a certain level of trust in its professionalism and reliability. [Scale]

(1 = Strongly Disagree, 5 = Strongly Agree)

○ 1  ○ 2  ○ 3  ○ 4  ○ 5

**Q7** ChatGPT has high professionalism and reliability and can effectively complete tasks. [Scale]

(1 = Strongly Disagree, 5 = Strongly Agree)

○ 1  ○ 2  ○ 3  ○ 4  ○ 5

**Q8** Based on ChatGPT's past performance, I believe it is good at handling complex tasks. [Scale]

(1 = Strongly Disagree, 5 = Strongly Agree)

○ 1  ○ 2  ○ 3  ○ 4  ○ 5

**Q9** In actual use, ChatGPT operates accurately and reliably, with few technical issues disrupting my workflow. [Scale]

(1 = Strongly Disagree, 5 = Strongly Agree)

○ 1  ○ 2  ○ 3  ○ 4  ○ 5

**Q10** My friends and colleagues generally believe that ChatGPT has high professionalism and credibility. [Scale]

(1 = Strongly Disagree, 5 = Strongly Agree)

○ 1 ○ 2 ○ 3 ○ 4 ○ 5

**Q11** As I gain a deeper understanding of AI technology principles and background, I pay more attention to ChatGPT's performance. [Scale]

(1 = Strongly Disagree, 5 = Strongly Agree)

○ 1 ○ 2 ○ 3 ○ 4 ○ 5

**(Section 4) Institutional Trust**

**Q1** What is your level of trust in the Central Government? [Scale]

(1 = Completely Untrustworthy, 5 = Completely Trustworthy)

○ 1 ○ 2 ○ 3 ○ 4 ○ 5